# Molecular dynamics simulation study of mechanical properties of 3C-SiC with extended defects

Serhii Shmahlii[1] and Andrey Sarikov[1-3]

[1] V. Lashkaryov Institute of Semiconductor Physics, National Academy of Sciences of Ukraine, 41 Nauky Avenue, 03028 Kyiv, Ukraine

[2] Educational Scientific Institute of High Technologies, Taras Shevchenko National University of Kyiv, 4-g Hlushkova Avenue, 03022 Kyiv, Ukraine

[3] National Technical University of Ukraine "Igor Sikorsky Kyiv Polytechnic Institute", 37 Beresteiskyi Avenue, 03056 Kyiv, Ukraine

Corresponding author's email: sarikov@isp.kiev.ua

**Abstract**

In this study, large-scale molecular dynamics simulations with the Vashishta potential and the analytic bond-order potential (ABOP) were performed to investigate the effect of extended defects on the elastic properties of cubic silicon carbide (3C-SiC). Specifically, we focused on systems containing Shockley partial dislocations terminating stacking faults, along with double and triple dislocation complexes. The changes in the independent elastic stiffness constants $C_{11}$, $C_{12}$ and $C_{44}$ upon varying the mentioned extended defects concentrations were quantified. Using the values of these constants, the effective bulk, shear, and Young's moduli were calculated for different defect types and concentrations. The moduli were calculated along particular crystallographic directions aligned with the mentioned defect configurations as well as evaluated using Voigt-Reuss-Hill averaging to provide overall orientation-independent characterization of the defect-altered lattice. The obtained results reveal a general trend of diminishing the material's stiffness with increasing densities of Shockley partial dislocations and dislocation complexes. Depending on the defect configuration, the average elastic moduli decrease by up to approximately 6 % with the Vashishta potential and up to about 4 % using the analytic bond-order potential. At this, triple dislocation complexes induce smaller perturbations. These findings demonstrate that extended defect networks can measurably modify the elastic response of 3C-SiC and should be considered in further scientific research and practical applications of this material.

## 1. Introduction

Si carbide is one of the new generation's semiconductor materials that has over 250 polytypes, only several of them being suitable for commercial fabrication [1]. Among them, cubic SiC polytype (3C-SiC) is characterized by a combination of mechanical, electrical, thermal and chemical properties that may be interesting for many device applications. Notably, 3C-SiC is the only SiC polytype that can be economically grown on Si substrates, which opens up prospects for its integration with existing Si technologies [2]. This material is one of the stiffest known semiconductors, demonstrating exceptional hardness, excellent wear and oxidation resistance, and resistance to acids – the qualities that make it highly suitable for micro-electromechanical systems, harsh environment sensors, and protective layers [3]. Due to its cubic symmetry, 3C-SiC is more elastically isotropic than hexagonal SiC polytypes such as 4H- and 6H-SiC. As a result, crack propagation in 3C-SiC is expected to exhibit less pronounced orientation dependence compared to hexagonal structures, where fracture behavior is strongly influenced by crystallographic direction. [4].

The most widespread and standard method of 3C-SiC film fabrication is the epitaxial growth of 3C-SiC on Si substrates. However, this approach introduces high densities of extended structural defects such as dislocations and stacking faults caused by the ~20 % lattice mismatch and the differences in the thermal expansion coefficients between the 3C-SiC and Si [5, 6]. Using cross-sectional transmission electron microscopy, Mendez et al. [7] observed linear misfit dislocations at the 3C-SiC/Si interface, mostly of the Lömer (90°) type, with one dislocation every five {111} planes. These dislocations alone cannot relieve the residual strain, which is further relaxed by generation of perfect dislocations splitting into 30° and 90° Shockley partial dislocations (SPDs) separated by stacking faults [8, 9]. The 90° partials are largely confined to the interface, while 30° dislocations dominate the internal defect structure of the film through defect propagation and interaction [10, 11]. High-resolution transmission electron microscopy also revealed complex dislocation structures at double and triple stacking fault boundaries [12].

Experimentally, there are practically no ways to fabricate 3C-SiC films with a particular density of required types of extended defects without introducing other defect types. Therefore, experimental investigations of influence of particular defects on 3C-SiC characteristics can hardly be performed. Instead, theoretical approaches such as molecular dynamics (MD) and density functional theory (DFT) simulations can enable researchers to gain insights into the mentioned issue.

In [13], the structures of partial dislocation complexes in 3C-SiC were identified by molecular dynamics simulations. It was demonstrated that partial dislocations with identical Burgers vectors repulse each other. In contrast, stable configurations (extrinsic partial dislocations) are formed by 30° and 90° SPDs, or two 90° SPDs with opposite screw components of the Burgers vector, as well as three SPDs forming a net-zero Burgers vector triple dislocation complex. These complexes were found to introduce electronic states within the 3C-SiC band gap being, therefore, “killer defects” detrimental to the material electronic performance [14].

The results of the MD simulations [15] demonstrate that the effective melting temperature of 3C-SiC tends to decrease with an increase of the concentration of single partial dislocations, eventually levelling off at around 165 K below that of a perfect crystal. The extrinsic partial dislocations lower the melting point by only about 50 K regardless of their concentration, while the impact of triple dislocation complexes on the melting temperature is negligible.

MD simulations of the SiC’s mechanical properties using Tersoff-type potentials [16] claim that point defects such as vacancies, interstitials, and antisite defects, as well as their small clusters, have a much stronger degrading impact on mechanical properties than large-scale structural disruptions in the atomic arrangement of the crystalline material, such as amorphous pockets, dislocation networks, etc. The study based on the DFT method [17] reveals that point defects in SiC generally reduce the elastic moduli. The most significant effect is observed with carbon vacancies, which lead to a drop in the Young’s modulus to 234 GPa and the shear modulus to 89.6 GPa — both by about 50 % lower than the values of 433 and 187 GPa for ideal crystal, respectively. Si vacancies and interstitials also notably weaken the material, reducing the bulk, Young’s, and shear moduli by approximately 9.9-10.7 %, 18.9-21 %, and 21.2-23.6 %, respectively. Besides, SiC antisite defects cause a 12.4 % decrease in the bulk modulus.

However, stacking faults terminated by Shockley partial dislocations as well as double and triple dislocation complexes raise still unanswered questions about their effects on the mechanical properties of cubic silicon carbide. In this study, we use molecular dynamics simulations to examine how the mentioned extended defects at varying densities affect these properties. We choose classical molecular dynamics for investigations, first of all, for its ability to flexibly determine the elasticity tensor of large atomic ensembles, which is hardly realizable by either in situ experimental methods or DFT calculations for large systems needed to accommodate extended defects.

## 2. Method

### 2.1. Cell preparation and simulation method

In this work, we used an ideal 3C-SiC initial simulation cell with the dimensions of approximately 171 × 9 × 181 Å$^3$ containing 27648 atoms. The directions of the cell axes were $[1\overline{1}\overline{2}]$, [110] and $[1\bar{1}1]$. Such cell orientation was chosen to enable periodic boundary conditions on it in view of the preferential dislocation lines directions along <110>. The ideal cell was modified by inserting dipoles consisting of pairs of Shockley partial dislocations and dislocation complexes, as described in [15], to study the effect of typical of 3C-SiC extended defects on the material elastic properties. The dislocations were inserted by shifting all the atoms of the simulation cells by the vectors with the components determined in the framework of the dislocation theory as follows [18]:

$$u_x = \frac{b_{edge}}{2\pi}\left(tan^{-1}\frac{y}{x} + \frac{xy}{2(1-\nu)(x^2+y^2)}\right) \quad (1)$$

$$u_y = \frac{b_{edge}}{2\pi}\left(\frac{1-2\upsilon}{4(1-\upsilon)}\ln(x^2+y^2) + \frac{x^2-y^2}{4(1-\upsilon)(x^2+y^2)}\right) \quad (2)$$

$$u_z = \frac{b_{screw}}{2\pi}tan^{-1}\frac{y}{x} \quad (3)$$

where $b_{edge}$ and $b_{screw}$ are the edge and screw components of the dislocation Burgers vector, respectively, and ν is the Poisson ratio of 3C-SiC. The cells with 2, 4, 6 and 8 dipoles formed by pairs of 90° and 30° Shockley partial dislocations with stacking faults in between, 1, 2, 3, and 4 extrinsic partial dislocation (EPD) dipoles, and 1 and 2 triple dislocation complex dipoles were created [15]. The dislocation dipoles were inserted so to ensure zero total Burgers vector of the cell. Periodic boundary conditions were applied to the simulation cells in all the three directions to mimic bulk 3C-SiC material behavior.

Molecular dynamics simulations were carried out with a Large-scale Atomic/Molecular Massively Parallel Simulator (LAMMPS) software [19] using Vashishta and analytic bond-order potential (ABOP) to account for interaction of Si and C atoms. These potentials were employed due to their ability to capture various effects of extended defect structures and behavior in cubic SiC, while remaining computationally efficient for large-scale atomistic simulations [20]. Visualization and analysis of the obtained cell configurations were carried out using an Open Visualization Tool (OVITO) software [21]. The percentages of the extended defects and ensued hexagonal diamond polytype structure in all the employed simulation cells are given in Table 1.

Table 1. Percentage of extended defects and hexagonal diamond polytype structure in the 3C-SiC cells

| Defects configurations in 3C-SiC cell | Percentage (%) of defects in 3C-SiC cell | Percentage (%) of hexagonal diamond structure in 3C-SiC cell |
|---|---|---|
| ideal | 0 | 0 |
| 2 pairs SPDs | 0.9 | 2.4 |
| 4 pairs SPDs | 2.3 | 4.9 |
| 6 pairs SPDs | 3.5 | 7.5 |
| 8 pairs SPDs | 4.1 | 10.2 |
| 1 pair EPD | 0.9 | 1.3 |
| 2 pairs EPDs | 2.6 | 2.6 |
| 3 pairs EPDs | 3.5 | 4.1 |
| 4 pairs EPDs | 5.8 | 5.5 |
| 1 pair of triple complexes | 1.4 | 1.4 |
| 1 pair of triple complexes | 1.6 | 1.6 |

**2.2. Calculation of stiffness tensor components**

The stiffness tensor indicates how an anisotropic material deforms under stress in different directions, which is crucial for assessing the material's mechanical characteristics. In this work, the stiffness tensor components were calculated by applying finite volume deformations to the simulation cells using the adapted script included in the LAMMPS installation package.

The key elasticity equation in the Einstein notation [22] is

$$\sigma_{ij} = C_{ijkl}\varepsilon_{kl} \tag{4}$$

where $\sigma_{ij}$ and $\varepsilon_{kl}$ are the stress and strain components, respectively, and the stiffness tensor $C_{ijkl}$ has the symmetries $C_{ijkl} = C_{klij} = C_{jikl} = C_{ijlk}$ [23].

Ideal 3C-SiC material has a cubic crystal symmetry, significantly simplifying its stiffness tensor $C_{ij,}$ which in Voigt notation (6×6 matrix) typically includes only three independent components $C_{11}$, $C_{12}$, and $C_{44}$, and generally looks as follows [24, 25]:

$$C = \begin{bmatrix} C_{11} & C_{12} & C_{12} & 0 & 0 & 0 \\ C_{12} & C_{11} & C_{12} & 0 & 0 & 0 \\ C_{12} & C_{12} & C_{11} & 0 & 0 & 0 \\ 0 & 0 & 0 & C_{44} & 0 & 0 \\ 0 & 0 & 0 & 0 & C_{44} & 0 \\ 0 & 0 & 0 & 0 & 0 & C_{44} \end{bmatrix} \quad (5)$$

Introducing defects into the 3C-SiC crystal structure disrupts the cubic symmetry and can typically manifest itself by non-zero values in the places of zeros in expression (5) as well as by violation of the equalities $C_{11} = C_{22} = C_{33}$, $C_{12} = C_{13} = C_{23}$, and $C_{44} = C_{55} = C_{66}$. As a result, the stiffness tensor might get a more complicated look:

$$C = \begin{bmatrix} C_{11} & C_{12} & C_{13} & C_{14} & C_{15} & C_{16} \\ C_{12} & C_{22} & C_{23} & C_{24} & C_{25} & C_{26} \\ C_{13} & C_{23} & C_{33} & C_{34} & C_{35} & C_{36} \\ C_{14} & C_{24} & C_{34} & C_{44} & C_{45} & C_{46} \\ C_{15} & C_{25} & C_{35} & C_{45} & C_{55} & C_{56} \\ C_{16} & C_{26} & C_{36} & C_{46} & C_{56} & C_{66} \end{bmatrix} \quad (6)$$

If the deviation from the above-mentioned equalities is insignificant, the following averaging may be used [26]:

$$\overline{C}_{11} = \frac{C_{11}+C_{22}+C_{33}}{3} \quad (7)$$

$$\overline{C}_{12} = \frac{C_{12}+C_{13}+C_{23}}{3} \quad (8)$$

$$\overline{C}_{44} = \frac{C_{44}+C_{55}+C_{66}}{3} \quad (9)$$

Using the calculation method described above, we obtained the stiffness tensor components for the cells defined by the axes $X'[1\overline{1}\overline{2}]$, $Y'[110]$, and $Z'[1\overline{1}1]$. To obtain these components for the standard orientation of the rectangular coordinate system, namely defined by the axes $X[100]$, $Y[010]$ and $Z[001]$, an appropriate conversion procedure is required. First, we need to find the Euler's angles of rotation from the original coordinate system to the standard one, by composing 3×3 matrix with coordinates of the directional vectors of the $X'$, $Y'$ and $Z'$ axes as its columns and inverting it to form an inverse rotation matrix $R=\{R_{ij}\}$. The Euler angles α, β and γ are calculated using the elements of the matrix $R$ as follows:

$$\alpha = tan^{-1}\left(\frac{R_{32}}{-R_{12}}\right), \qquad \beta = cos^{-1}(R_{22}), \qquad \gamma = tan^{-1}\left(\frac{R_{23}}{R_{21}}\right) \tag{10}$$

The rotation matrices around the *X*[100], *Y*[010], and *Z*[001] axes of the standard coordinate system enabling transformation of the stiffness matrix elements from the initial system to systems with targeted orientations are given by

$$X = \begin{bmatrix} 1 & 0 & 0 & 0 & 0 & 0 \\ 0 & c^2 & s^2 & 2cs & 0 & 0 \\ 0 & s^2 & c^2 & -2cs & 0 & 0 \\ 0 & -cs & cs & c^2 - s^2 & 0 & 0 \\ 0 & 0 & 0 & 0 & c & -s \\ 0 & 0 & 0 & 0 & s & c \end{bmatrix} \tag{11}$$

$$Y = \begin{bmatrix} c^2 & 0 & s^2 & 0 & 2cs & 0 \\ 0 & 1 & 0 & 0 & 0 & 0 \\ s^2 & 0 & c^2 & 0 & -2cs & 0 \\ 0 & 0 & 0 & c & 0 & -s \\ -cs & 0 & cs & 0 & c^2 - s^2 & 0 \\ 0 & 0 & 0 & s & 0 & c \end{bmatrix} \tag{12}$$

$$Z = \begin{bmatrix} c^2 & s^2 & 0 & 0 & 0 & 2cs \\ s^2 & c^2 & 0 & 0 & 0 & -2cs \\ 0 & 0 & 1 & 0 & 0 & 0 \\ 0 & 0 & 0 & c & s & 0 \\ 0 & 0 & 0 & -s & c & 0 \\ -cs & cs & 0 & 0 & 0 & c^2 - s^2 \end{bmatrix} \tag{13}$$

where $c$ is the cosine and $s$ is the sine of the above-mentioned Euler angles α, β and γ. The rotation matrix for transformation of our stiffness matrix elements to those in the standard coordinate system is

$$R = Y_\alpha Z_\beta Y_\gamma. \tag{14}$$

Here, the lower indices indicate the angles put in every used transformation matrix. We can get the stiffness matrix for the standard orientation $C_{st}$ as follows:

$$C_{st} = R \times C \tag{15}$$

where $C$ is the stiffness matrix in the original coordinate system.

## 2.2 Calculation of Young's and shear moduli along crystallographic directions aligned with stacking fault–Shockley partial dislocation geometry in 3C-SiC

The elastic moduli were evaluated along the crystallographic directions linked to the stacking fault–Shockley partial dislocation geometry in 3C-SiC. The directional Young's moduli were obtained by contracting the full anisotropic compliance tensor with the unit vectors aligned normal to the stacking fault plane, parallel to the Shockley partial dislocation line, and in-plane perpendicular to the dislocation line [26]:

$$E^{-1} = S_{ijkl} d_j d_k d_l \tag{16}$$

Here, $S$ is the compliance tensor, which can be computed as the inverse of the stiffness tensor, $E$ is the Young's modulus, and $d_i$ are the direction cosines of the load axis, respectively. To obtain an expression for calculations, it is possible to take advantage of the above-mentioned fact that the compliance tensor $S_{ijkl}$ has the symmetries $S_{ijkl} = S_{klij} = S_{jikl} = S_{ijlk}$, which enables conversion from the 4th-order tensor notation to the Voigt notation (6×6 matrix form) and ultimately get the following equation:

$$E(d) = (S_{11}d_1^4 + S_{22}d_2^4 + S_{33}d_3^4 + +(2S_{12} + S_{66})d_1^2 d_2^2$$
$$+(2S_{13} + S_{55})d_1^2 d_3^2 + (2S_{23} + S_{44})d_2^2 d_3^2)^{-1} \tag{17}$$

The normal direction to the stacking fault plane is $[1\bar{1}1]$. It probes fault-normal stiffness and interplanar bonding. The direction [110] follows the partial dislocation line. The in-plane transverse direction $[1\bar{1}\bar{2}]$ orthogonal to [110] captures fault-plane elastic anisotropy.

The shear modulus associated with shear along the direction **m** in a plane with normal **n** was obtained from the compliance tensor as

$$\frac{1}{G(n,m)} = S_{ijkl} n_i m_j n_k m_l \tag{18}$$

where $n \times m = 0$, $n_i$ are the components of a unit vector in the direction of interest, and $m_i$ are the components of a unit vector perpendicular to $n_i$, respectively. Expressing the deformation in terms of symmetric strain components and adopting Voigt notation with engineering shear strains, the equation (18) reduces to the computationally efficient form:

$$G(n,m) = \frac{1}{4\sum_{i=1}^{6}\sum_{j=1}^{6}\epsilon_i S_{ij}\epsilon_j} \quad (19)$$

where $\epsilon_i$ are the components of the strain vector in row form and $\epsilon_j$ are the components of the strain vector in column form, respectively. The strain vector is

$$\epsilon = [n_1m_1, n_2m_2, n_3m_3, n_2m_3 + n_3m_2, n_1m_3 + n_3m_1, n_1m_2 + n_2m_1]^T \quad (20)$$

or

$$\epsilon = \begin{bmatrix} n_1m_1 \\ n_2m_2 \\ n_3m_3 \\ n_2m_3 + n_3m_2 \\ n_1m_3 + n_3m_1 \\ n_1m_2 + n_2m_1 \end{bmatrix} \begin{matrix} \leftarrow \{contribution\ of\ shear\ to\ normal\ strain\ \}x \\ \leftarrow \{contribution\ of\ shear\ to\ normal\ strain\ \}y \\ \leftarrow \{contribution\ of\ shear\ to\ normal\ strain\ \}z \\ \leftarrow \{Shear\ \}yz \\ \leftarrow \{Shear\ \}xz \\ \leftarrow \{Shear\ \}xy \end{matrix} \quad (21)$$

where the first three components correspond to the contribution of shear to normal strain along the *X*, *Y* and *Z* axes and the last three components correspond to shear strain in the *YZ*, *XZ* and *XY* planes. In this notation, $n_i$ stands for unit normal vector to the plane, and $m_j$ stands for the direction within that plane. This construction ensures that the evaluated shear modulus corresponds to deformation along a physically meaningful crystallographic slip system.

The shear moduli were computed along physically relevant directions: in-plane glide direction $(1\overline{1}1)[110]$ governing partial dislocation motion (deformation in the direction [110] within the $(1\overline{1}1)$ plane), transverse in-plane shear direction $(1\overline{1}1)[1\overline{1}\overline{2}]$ probing fault-plane anisotropy, and direction of shear normal to the stacking fault $[110]\perp[1\overline{1}1]$ capturing out-of-plane coupling. This set dictated by the cubic SiC crystallography allows a minimal yet complete characterization of the defect-induced elastic anisotropy.

### 2.3. Calculation of elastic constants

Elastic constants in tensor/matrix forms are direction-dependent and numerous, so it is expedient to obtain more aggregate characteristics of the material elastic properties. Using the computed full 6×6 stiffness matrix $C_{ij}$ from our simulations with LAMMPS, we can calculate the elasticity moduli resorting to the following Voigt-Reuss-Hill (VRH) approximation [27]. As the concentration of extended defects grows, the material behaves less like a single crystal and more like a quasi-isotropic medium. The VRH average accounts for this structural disorder by calculating a reliable 'middle-ground' value (Hill average) between the theoretical upper (Voigt) and lower (Reuss) elastic limits. This approach allows for a direct comparison between the fundamental elastic constants of the lattice and the macroscopic mechanical behavior observed in real, defective samples. Although Voigt-Reuss-Hill averaging was initially designed for finding the elastic moduli of polycrystalline materials, further studies demonstrated its applicability for single crystalline materials [28]. The work [17] investigates the effect of defects on the elastic moduli of single crystalline SiC, employing the Voigt-Reuss-Hill technique. Therefore, we can use the same equations in our study of the impact of extended defects on the 3C-SiC mechanical characteristics.

The Voigt and Reuss averages of the bulk modulus:

$$B_V = B_R = B = \frac{1}{3}(C_{11} + 2C_{12}) \tag{22}$$

The Voigt $G_V$, Reuss $G_R$, and Hill $G_H$ averages of the shear modulus are calculated as follows [17, 27, 28]:

$$G_V = \frac{1}{5}(C_{11} - C_{12} + 3C_{44}) \tag{23}$$

$$G_R = \frac{5(C_{11} - C_{12})C_{44}}{4C_{44} + 3(C_{11} - C_{12})} \tag{24}$$

$$G_H = \frac{G_V + G_R}{2} \tag{25}$$

Using the average value of the bulk modulus and the Hill average of the shear modulus, the value of the Young's modulus $E$ can be found as

$$E = \frac{9BG_H}{3B + G_H} \tag{26}$$

Using the elastic stiffness tensor, one can also quantify how anisotropic a material is, based on its Universal Elastic Anisotropy Index ($A^U$), a scalar measure of elastic anisotropy introduced in [29], which for the cubic Si carbide looks as follows:

$$A^U = 5\left(\frac{G_V}{G_R} - 1\right) \tag{27}$$

Transverse wave velocity is the velocity at which shear (transverse) waves move through an elastic material [17]:

$$v_t = \sqrt{\frac{G}{\rho}} \tag{28}$$

where $G$ is the shear modulus and ρ is the mass density of the material, respectively.

The longitudinal wave velocity (also called P-wave velocity) is the speed at which compressional or pressure waves spread through a solid. These waves cause particles in the material to oscillate parallel to the direction of wave propagation [17]:

$$v_l = \sqrt{\frac{B+\frac{4}{3}G}{\rho}} \tag{29}$$

One may also consider these velocities as indicators of resistance to compression and shear deformations – the lower the velocities are, the higher the respective resistances are. The transverse and longitudinal wave velocities are used to find the average (mean) sound velocity [17]:

$$v_m = \left[\frac{1}{3}\left(\frac{2}{v_t^3} + \frac{1}{v_l^3}\right)\right]^{-1/3} \tag{30}$$

which is a crucial factor to compute the Debye temperature:

$$\Theta_D = \frac{h}{k_B}\left(\frac{3n}{4\pi V}\right)^{1/3} v_m = \frac{h}{k_B}\left[\frac{3n}{4\pi}\left(\frac{N_A \rho}{M}\right)\right]^{1/3} v_m \tag{31}$$

Here, $k_B$ is the Boltzmann constant, $h$ is the Planck’s constant, and $V$ is the atomic volume, respectively. The Debye temperature is a fundamental material characteristic that corresponds to the highest vibrational (phonon) frequency in a solid:

$$\Theta_D = \frac{h v_D}{k_B} \tag{32}$$

where $\nu_D$ is the Debye (maximum) phonon frequency and $\Theta_D$ sets the energy limit for phonons:

$$E_{max} = h\nu_D = k_B\Theta_D \quad (33)$$

At low temperatures ($T \ll \Theta_D$), only low-energy phonons are excited that is heat capacity drops sharply. At high temperatures ($T \gg \Theta_D$), all modes are excited and, consequently, there is classical behavior (Dulong–Petit law).

## 3. Results

### 3.1. Independent elastic constants

The independent elements of the elastic tensor for ideal 3C-SiC obtained from our MD simulations were compared with respective literature data (see Table 2). As can be seen from Table 2, MD simulations with the Vashishta potential and ABOP for ideal 3C-SiC provide physically viable values that are within the ranges previously reported in both theoretical and experimental studies. At this, ABOP ensures excellent consistency with the experimental data, while the Vashishta potential provides a good agreement for $C_{11}$ and $C_{12}$ but underestimates the $C_{44}$ value.

Table 2. Independent elastic constants: the results of the present work vs literature data

| Data source | $C_{11}$ (GPa) | $C_{12}$ (GPa) | $C_{44}$ (GPa) |
|---|---|---|---|
| Our MD results using Vashishta potential | 390.11 | 142.61 | 136.94 |
| Our MD results for ideal 3C-SiC using ABOP | 383.78 | 144.41 | 238.97 |
| Theoretical results [30][1] | 300–451 (352–420) | 101–214 (126–163) | 136–287 (211–287) |
| Experimental results [31] | 395 ± 12 | 132 ± 10 | 236 ± 7 |

[1] The ranges of all the reported theoretical values are provided (in parenthesis are the ranges of the values obtained by first-principles calculations)

Figure 1 presents the values of the independent elastic constants simulated using Vashishta potential and ABOP. As can be seen from Figure 1(a), the mentioned values obtained using the Vashishta potential tend to slightly decrease with an increase in the concentrations of each considered extended defect type. The maximum decrease by about 5.4 % for the $C_{11}$ value is observed for the cell with 4 pairs of external partial dislocations, followed by the decrease of $C_{44}$ by 5.1 %. As can be further inferred from Figure 1(a), the decrease in the $C_{12}$ value does not exceed about 1.5 %.

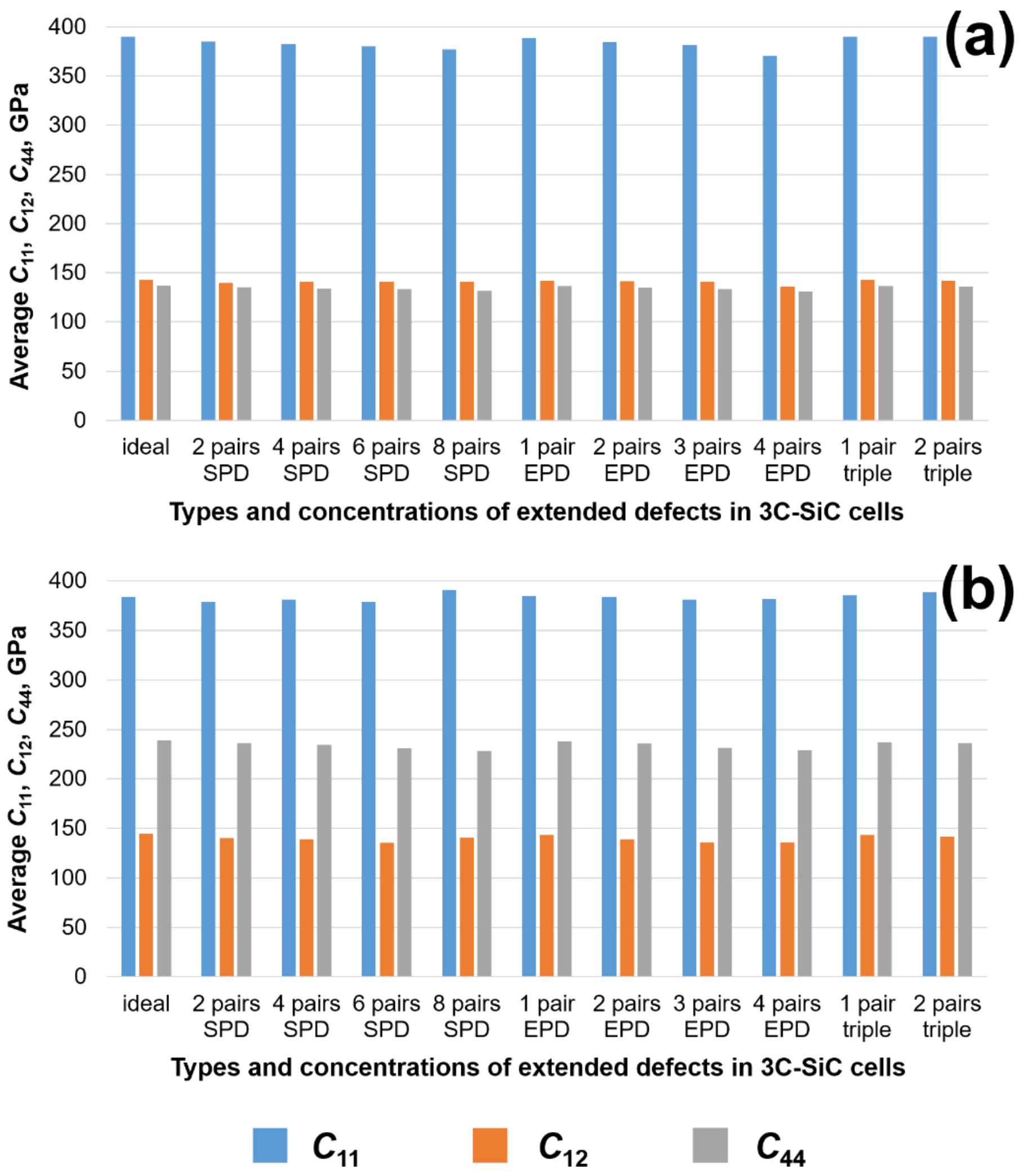

Figure 1. Influence of concentration and type of extended defects on independent elastic constants with (a) Vashishta potential and (b) ABOP.

As shown in Figure 1(b), the values of $C_{11}$ calculated with ABOP are generally less sensitive to the increase in the extended defect concentration compared to those obtained with Vashishta potential, decreasing by no more than 1.5 % as compared to the ideal cell value. External partial dislocations and triple dislocation complexes induce insignificant fluctuations on the $C_{11}$ values of about 1 %. $C_{12}$ generally goes down with an increase in the defect concentration. The maximal decrease in the $C_{12}$ value does not exceed 6.3 %, while this value for the cell with 8 pairs of dislocations roughly bounces back to the value of $C_{12}$ of the cell with 2 pairs of dislocations. The value of $C_{44}$ exhibits a steady decline by about 4.4 % for the cell with 8 pairs of dislocations and 4 pairs of external partial dislocations. The possible reason for both fluctuations and absence of a steady decrease in the values of the independent elastic constants with increase in the elastic defect concentrations is a higher ABOP's sensitivity to local effects such as strengthening interactions between densely periodically packed defects [32].

### 3.2 Young's and shear moduli along crystallographic directions linked to extended defect geometries

Figure 2(a) presents the Young' moduli for 3C-SiC with above mentioned defect configurations in the three direction: the direction $[1\bar{1}1]$ normal to the stacking fault plane that probes fault-normal stiffness and interplanar bonding, the direction [110] that lies within the $(1\bar{1}1)$ plane and corresponds to the dislocation line direction, and the in-plane transverse direction $[1\bar{1}\bar{2}]$, orthogonal to [110], which captures fault-plane elastic anisotropy.

The Young's moduli for the selected directions obtained using the Vashishta potential demonstrate up to 5.8 % decrease in the value in the third direction for the cell with 4 EPDs, followed by the 4.52 % decrease for the same defect configuration but in the first direction and up to 3.7 % decrease for the cell with 8 SPDs. One and two triple dislocation complexes induce less than 0.5 % decrease in the Young's modulus values for all the three directions. The rest of the cells show a decrease in the Young's modulus values, but within 1-3 % with respect to the ideal cell. The general trend for the results in the mentioned directions obtained with Vashishta potential is a consistent decrease in the Young's modulus with an increase in the extended defects concentration.

MD simulations with ABOP reveal a pronounced anisotropy of the first direction with respect to the second and the third ones, with overall higher values of the Young's modulus as compared to the Vashishta potential case. Rise in the concentration of the extended defects results in the strongest (3.9 %) decrease of the Young's modulus in the first direction for the cell with 4 EPDs as well as a decrease by 3.1 % for the cell with 3 EPDs, by 3.46 % for the cell with 8 SPDs, and by 3.44 % for the cell with 6 SPDs. For the rest of the cells, the decrease of the Young's modulus does not exceed 2 %.

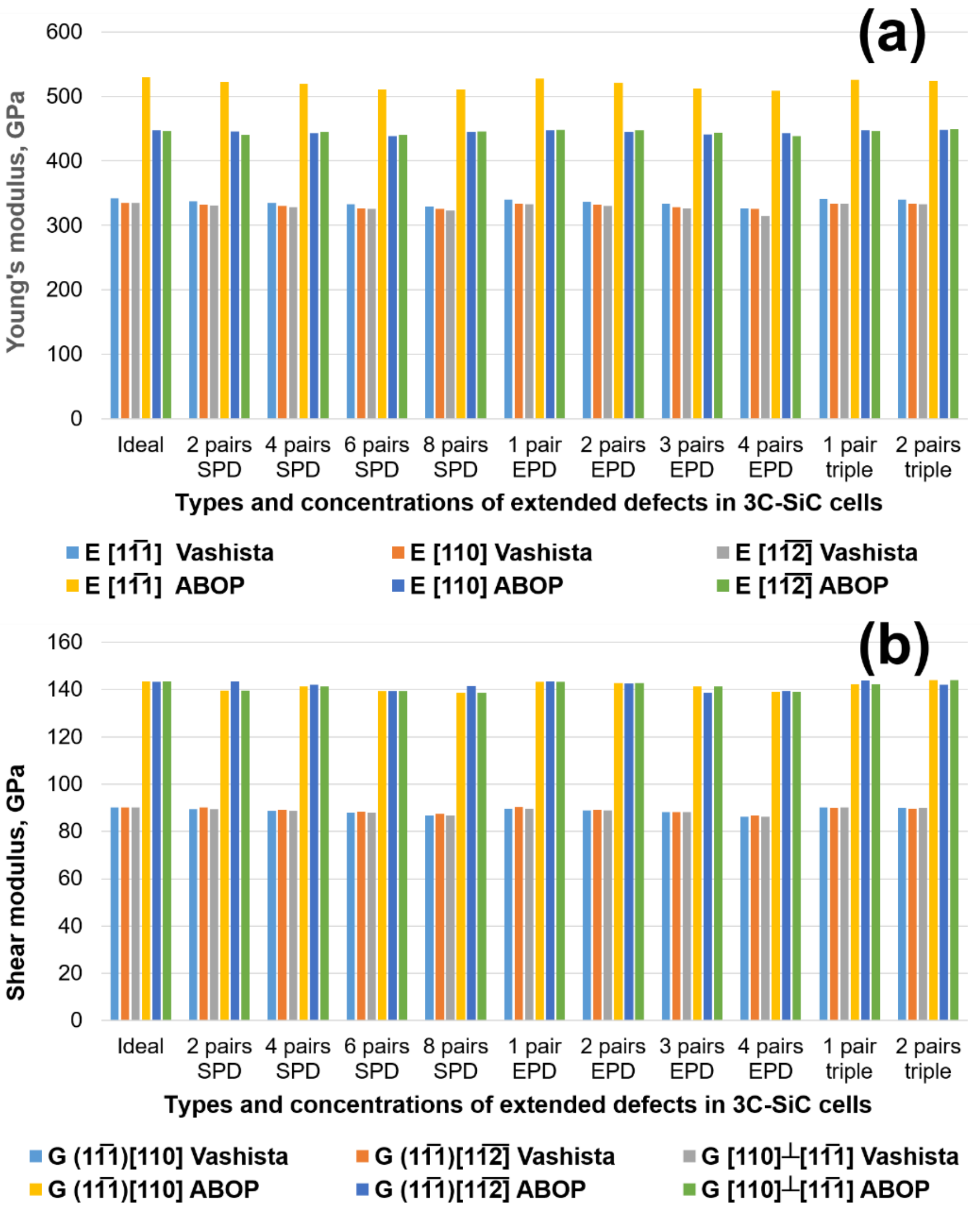

Figure 2. (a) Young's modulus *E* and (b) shear modulus *G* in directions defined by extended defect geometries.

According to Figure 2(b), the results of the MD simulations with Vashishta potential elicit that the shear modulus in the three directions linked to the 3C-SiC defect geometry decreases to the highest extent by up to 4.6-4.7 % for the cell with 4 EPDs relative to the value for the ideal cell. For the cell with 8 SPDs, the shear modulus decreases by 4 %, 3.3 % and 4 % in the three mentioned directions, respectively. The cells with all the other extended defect configurations exhibit less than 3.3 % decrease in the shear modulus values in these directions.

The shear moduli computed with ABOP are substantially larger, by about one third in absolute values than in the Vashishta potential case. Moreover, the data show a less steady decline trend at growing the defects concentrations. The maximum decrease in the shear modulus value amounts to about 3.6 %.

### 3.3 Anisotropy analysis

As Figure 3 shows, MD simulations with ABOP provide significantly higher anisotropy (0.45–0.6), while use of Vashishta potential results in almost isotropic 3C-SiC material behavior (0.01–0.015). This means that choice of potential strongly influences the predicted elastic response of the material in presence of defects. Increase of the defect concentrations causes the universal elastic anisotropy index to go up for the Vashishta potential and down for ABOP.

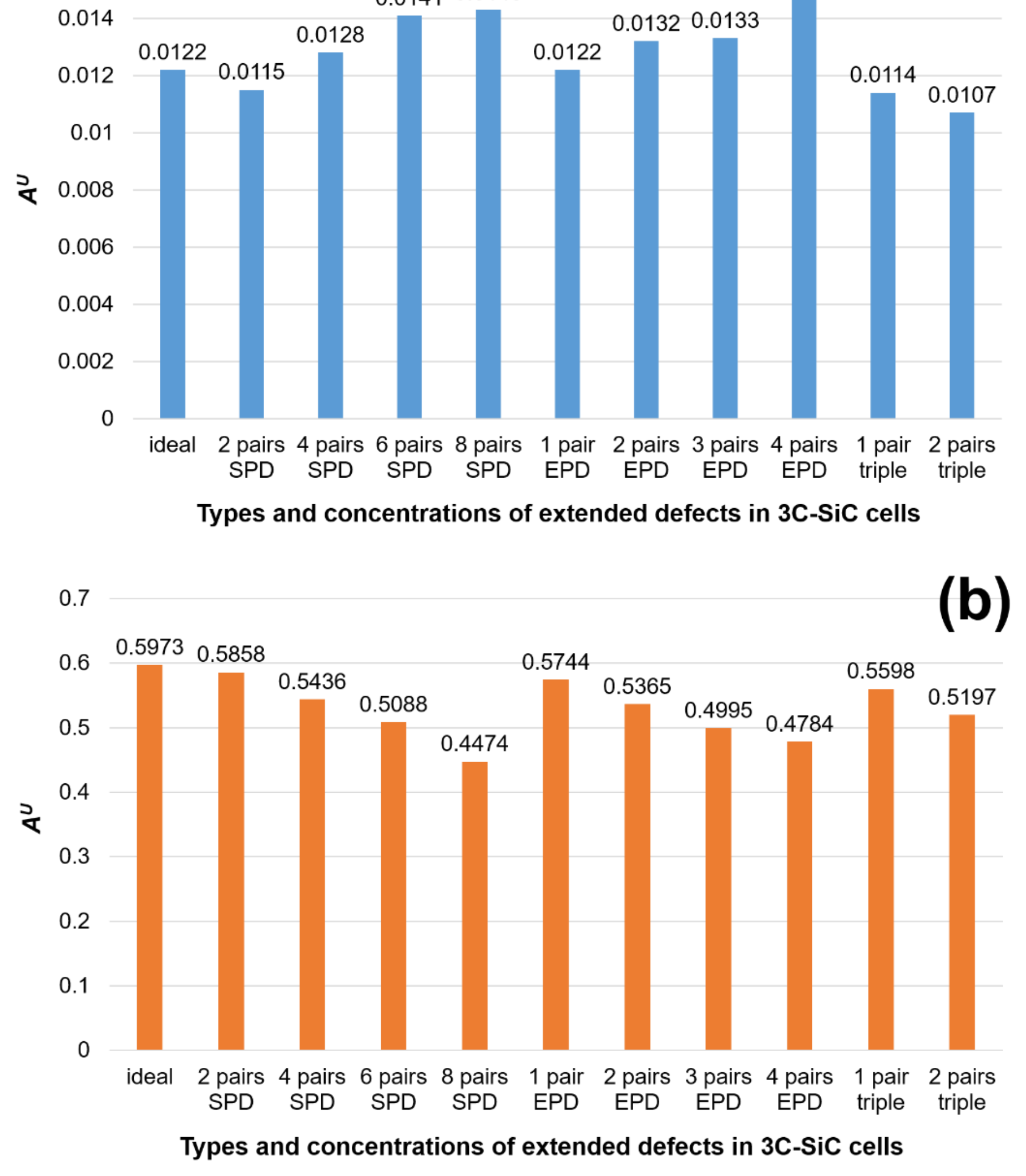

Figure 3. Universal elastic anisotropy index obtained with (a) Vashishta potential and (b) ABOP.

The universal elastic anisotropy index of isotropic materials is zero and positive deviations from zero mean more anisotropy. According to the experimental studies [33], mechanical response of single-crystalline 3C-SiC is inherently characterized by significant elastic anisotropy. The results of the MD simulations with Vashishta potential presented in Figure 3(a) indicate the anisotropy index by more than two orders of magnitude smaller than the value expected based on the experimental data. Therefore, modeling with this potential tends to sense only a vanishingly small enhancement of local elastic heterogeneity or anisotropic strain fields as the defect concentrations increase. Such poor accuracy is apparently due to the fact that Vashishta potential is parameterized based on cohesive energy and volumetric characteristics, making it well-suited to assess material's bulk properties, thermal characteristics, and structural transformations, at the same time seriously underestimating the independent elastic constant $C_{44}$ and thus causing poor anisotropy and shear moduli prediction.

In contrast, analytic bond-order potential provides reasonable values of the 3C-SiC elastic anisotropy index. By accurately capturing the component $C_{44}$, ABOP reproduces intrinsic directional stiffness of the cubic lattice, which makes this potential more reliable for simulating complex strain fields associated with stacking faults terminated by Shockley partial dislocations and their complexes. The decrease in the universal elastic anisotropy index from ~0.6 to ~0.45 at increasing density of extended defects indicates a progressive reduction in the elastic anisotropy of the material. This behavior may be attributed to defect-induced disruption of the crystalline lattice over extended regions, which tends to homogenize the directional dependence of the elastic response by introducing spatially distributed lattice distortions that overlap at higher defect densities. As a result, the contrast between different shear deformation modes is reduced, leading to a more uniform effective shear response and a decrease in elastic anisotropy.

### 3.4 Aggregate VRH elastic moduli

Figure 4 demonstrates that the VRH bulk moduli calculated with Vashishta potential and ABOP go down by no more than 5.1 % and 2.5 %, respectively, when the extended defect concentrations increase. Moreover, the bulk modulus found using ABOP tends to decrease less steadily at growing the extended defects density.

Figure 5(a) reveals that the VRH averaged Young's modulus obtained with the Vashishta potential tends to steadily decrease by up to 3.9 % and 4.8 % for the cells with the maximal concentrations of EPDs and SPDs respectively. The effect of triple dislocation complexes is negligibly small. The Young's modulus determined using ABOP goes down by up to 1.8 %, fluctuating in value at growing extended defects concentrations (see Figure 5(b)).

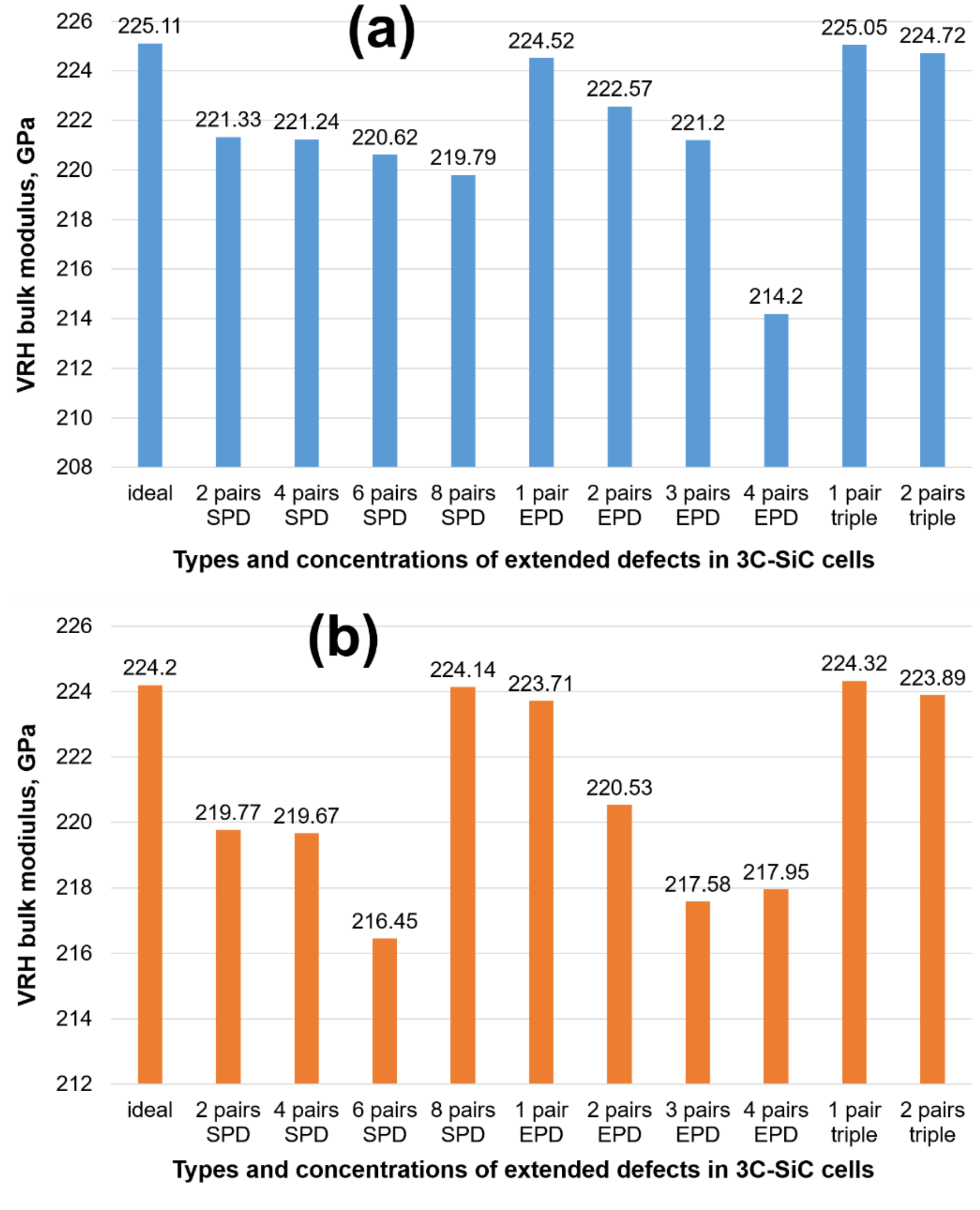

Figure 4. Effect of extended defects on VRH averaged bulk modulus obtained with (a) Vashishta potential and (b) ABOP.

Figure 6 shows the dependences of the VRH averaged shear moduli of 3C-SiC on the type and concentration of extended defects. Figure 6(a) illustrates that the shear modulus obtained with the Vashishta potential steadily goes down by up to 4.2 % and 5 % for the cell with 8 SPD pairs and 4 EPD pairs, respectively. In the case of ABOP (see Figure 6(b)), increase in the concentrations of EPDs and SPDs also leads to a decline in the shear modulus but by no more than 1.5 % with some fluctuations in the values for the cells with SPDs as well as triple dislocation complexes.

Figure 7 shows that increasing the defect density leads to a steady reduction in the Debye temperature $\Theta_D$. This trend is more pronounced for the Vashishta potential. The drop in the $\Theta_D$ value for the cell with 4 pairs of EPDs is 2.6 %. In its turn, simulations with ABOP yield an oscillatory decline of the mentioned characteristic with the span of about one per cent. At this, however, the absolute values of the Debye temperature determined with ABOP better correspond to the respective experimental and theoretical results [34].

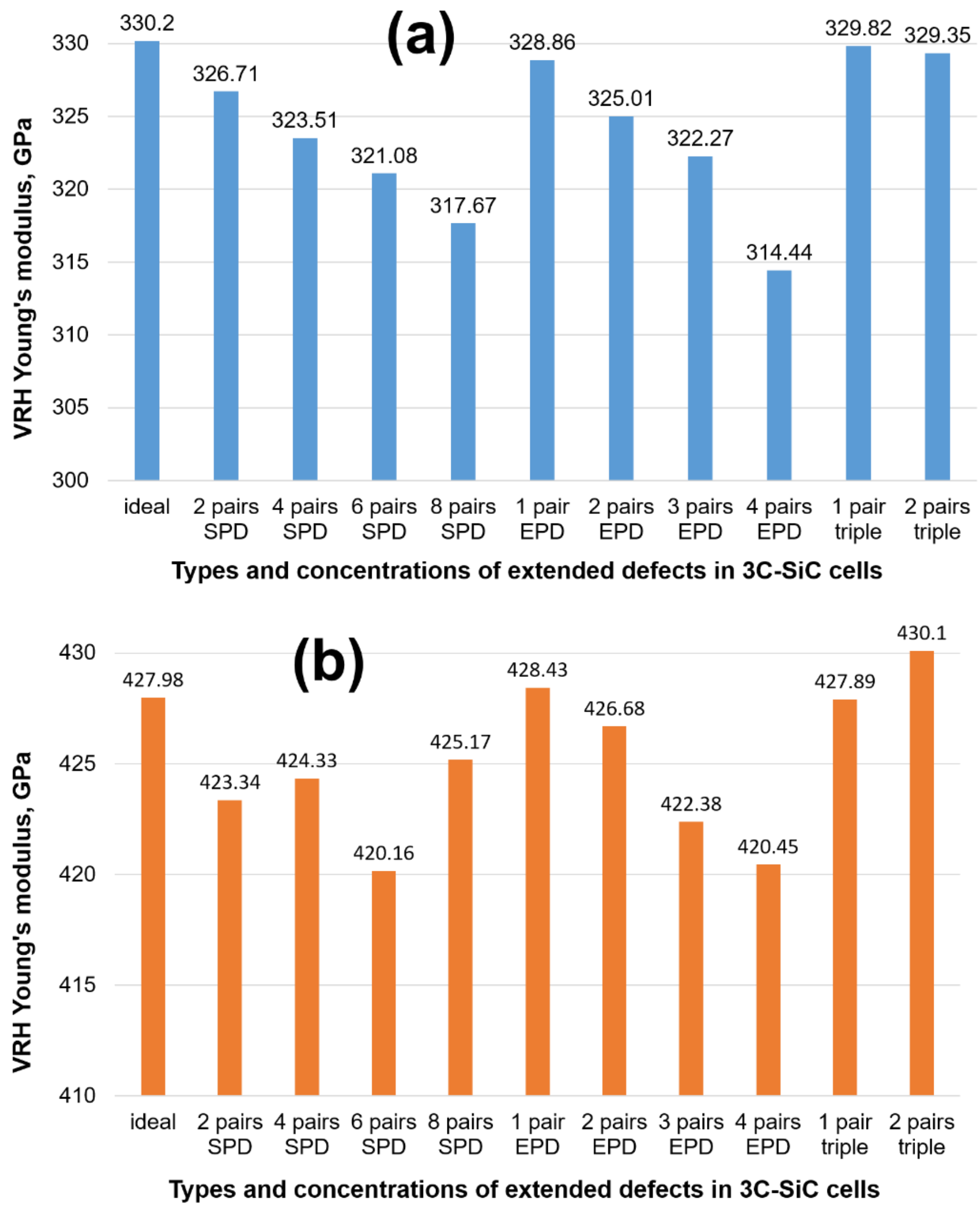

Figure 5. Effect of extended defects on VRH averaged Young's modulus obtained with (a) Vashishta potential and (b) ABOP.

## 4. Discussion

We calculated directional elastic moduli of cubic SiC containing extended defects using Vashishta and analytic bond-order interatomic potentials. The Young's modulus predicted with the Vashishta potential ranges between 314-342 GPa depending on the crystallographic direction, extended defect type, and their concentration. The directional shear modulus obtained with the Vashishta potential is approximately 87-90 GPa. In contrast, ABOP predicts significantly higher material stiffness, with the directional Young's modulus varying between 437-529 GPa and shear modulus around 138-143 GPa.

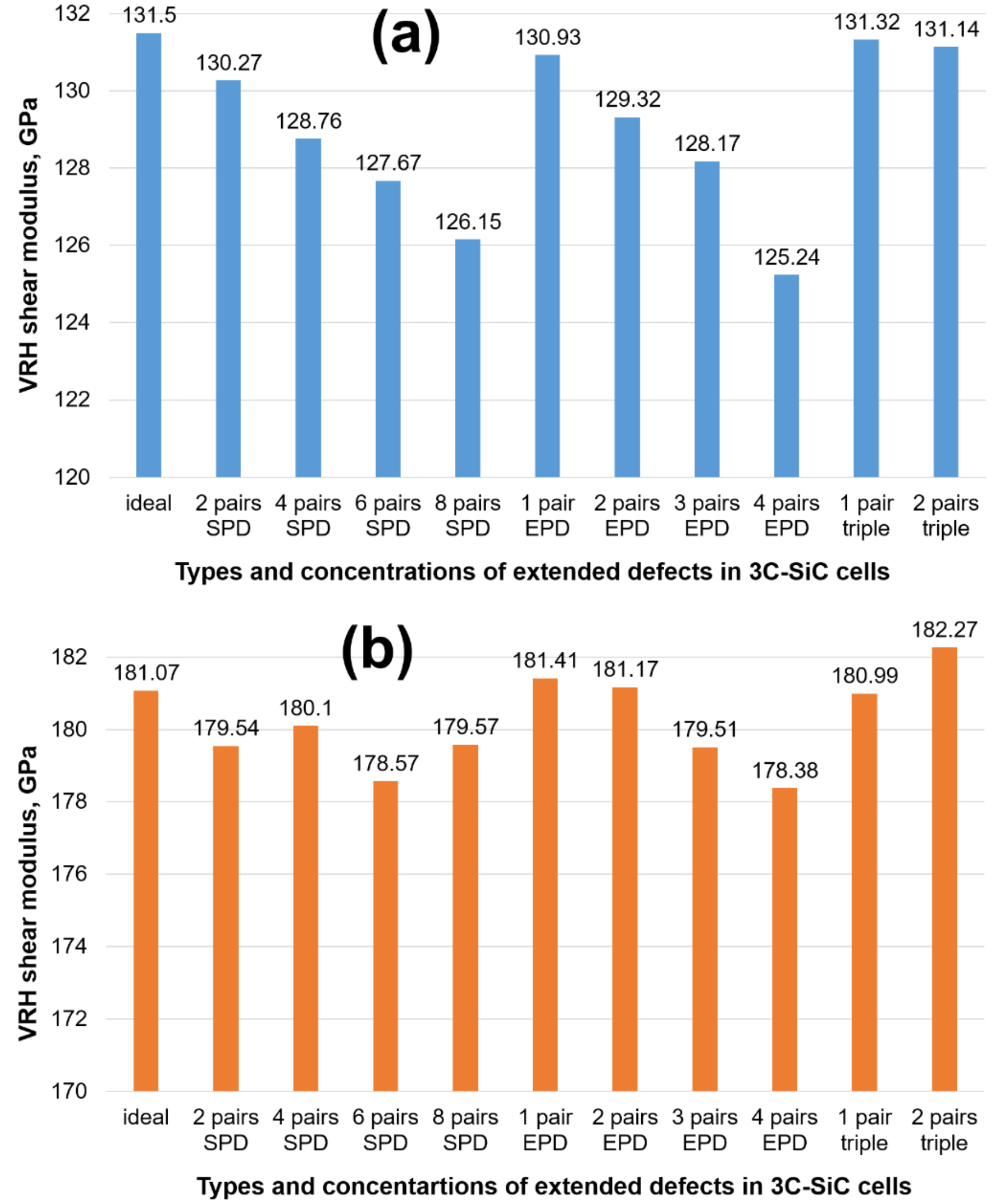

Figure 6. Effect of extended defects on VRH averaged shear modulus obtained with (a) Vashishta potential and (b) ABOP.

For different defect configurations, the VRH averaged Young's modulus ranges between 314-330 GPa for the Vashishta potential and between 420-430 GPa for ABOP. The VRH averaged values of the shear modulus lie within 125-131 GPa and 178-181 GPa obtained with Vashishta potential and ABOP, respectively. These results are summarized in Table 3.

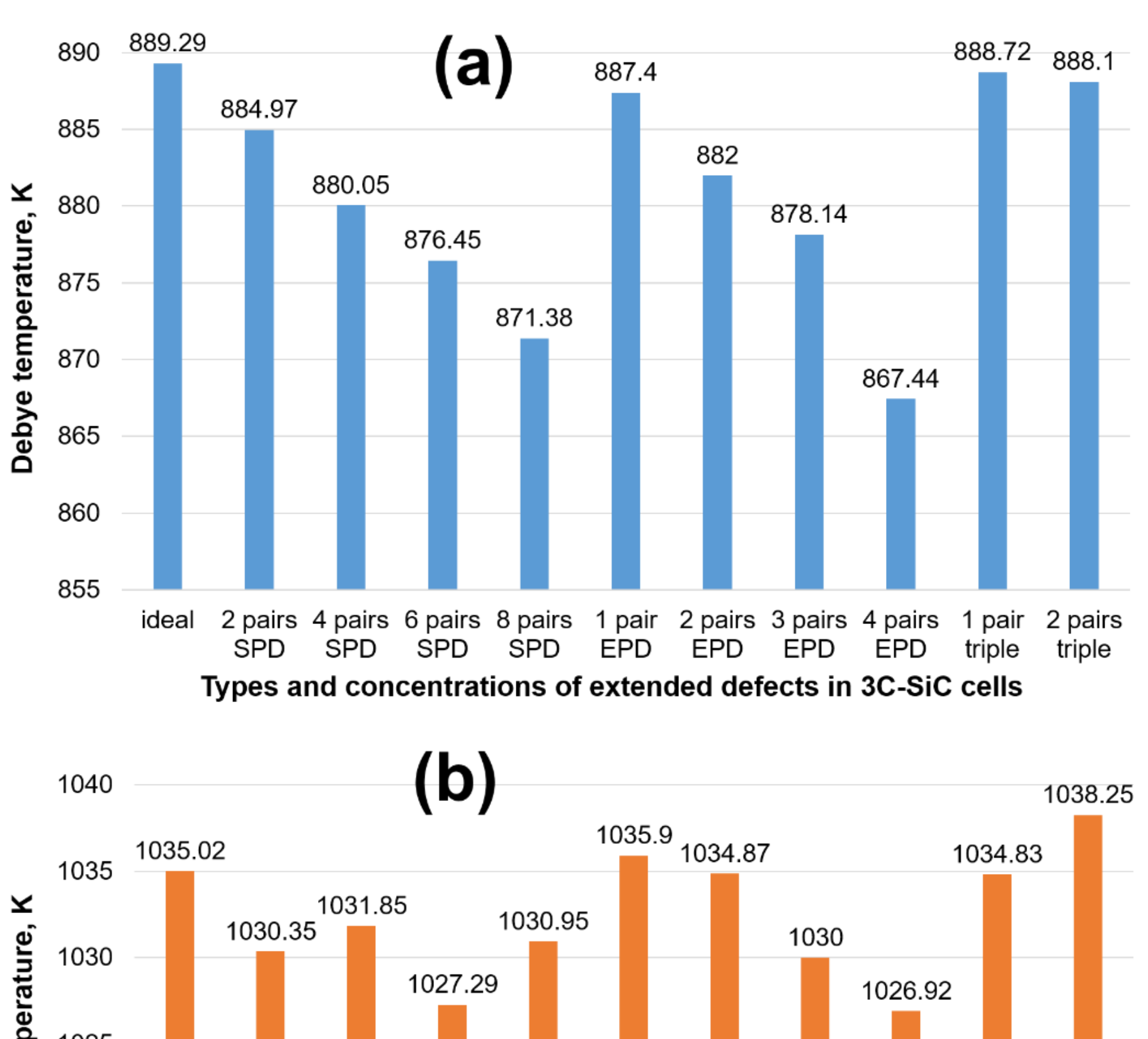

Figure 7. Effect of extended defects on the Debye temperature of 3C-SiC obtained with (a) Vashishta potential and (b) ABOP.

For single-crystalline or high-quality epitaxial 3C-SiC, the experimental room-temperature Young's modulus typically lies in the range of 400-450 GPa, as reported for high-quality films and micropillar structures [39, 41, 43]. The corresponding shear modulus is generally of the order of ~180-220 GPa, consistent with experimentally measured elastic constants.

In defected, irradiated, or polycrystalline thin-film 3C-SiC samples, such as those grown by chemical vapor deposition for microelectromechanical systems applications, the Young's modulus is often reduced to approximately 300-400 GPa [35, 36, 40, 42, 44]. This reduction is commonly associated with microstructural factors such as defects, grain boundaries, and residual stresses, which also lead to corresponding decrease in the effective shear modulus to approximately ~150-180 GPa.

Therefore, comparison of the MD simulated and experimental elastic moduli confirms that the MD simulations with ABOP provide better agreement with experiment than those with the Vashishta potential. As noted above, this discrepancy is mainly caused by the features of parametrization of the Vashishta potential, which prioritizes cohesive energy and volumetric material characteristics. While such parametrization makes this potential well-suited for assessing bulk properties, thermal behavior, and structural transformations, it results in a significant underestimation of the independent elastic constant $C_{44}$. Consequently, the potential yields less accurate predictions for elastic shear moduli, particularly in directions critical to dislocation mobility.

Table 3. VRH averaged and direction depended Young's and shear moduli vs literature experimental and theoretical data

| Methods | Young's modulus (GPa) | Shear modulus (GPa) |
|---|---|---|
| Experimental data [35-44][1] | 300-450 (400-450) | 150-220 (180-220) |
| Our results of VRH average with Vashishta potential | 314-330 | 125-131 |
| Our results of VRH average with ABOP | 420-430 | 178-181 |
| Our results for specific directions linked to the extended defects geometry obtained with Vashishta potential | 314-342 | 87-90 |
| Our results for specific directions linked to the extended defects geometry obtained with ABOP | 438-529 | 138-143 |

[1] The ranges of relevant 3C-SiC experimental values are provided (in parentheses are the ranges of the values obtained for high quality single crystal 3C-SiC)

The discrepancy between the experimental and VRH averaged values of the shear moduli of about 13 % and the MD assessment of the shear moduli in the specific directions related to the extended defect geometries is likely to be a classic example of scale-dependent mechanical response of the material. Experimentally, the shear modulus is typically measured across a macroscopic volume using acoustic waves or indentation, which probe statistical average over the entire sample. Because these samples consist of high-stiffness crystal domains separated by only thin planar defects, bulk measurements are dominated by the rigid crystal matrix, effectively "averaging out" localized softness provided by the defects. Furthermore, inherent background defects in experimental 3C-SiC samples may further mask the specific softening caused by extended defects. In contrast, MD simulations isolate the directional response in the specific defect-linked directions of the crystal, such as those associated with stacking faults and Shockley partial dislocations. MD simulations capture localized distortions of the $sp^3$ bonding and its transformation into a more compliant, hexagonal-like (2H) arrangement, providing a high-resolution view of the lattice softening that measurements of bulk samples cannot detect.

## 5. Conclusions

Our study reveals that increasing the concentration of extended defects – specifically Shockley partial dislocations terminating stacking faults as well as their double and triple complexes – generally degrades the mechanical properties of 3C-SiC, as evidenced by the decreased values of the elastic constants $C_{11}$, $C_{12}$, $C_{44}$, bulk, Young's, and shear moduli. The maximum decline of these values relative to the ideal crystal is about 5-6 % for the most extreme concentrations of the extended defects, with a more typical decrease in the range of 1-3 %. Triple complexes tend to affect the mechanical properties less than single Shockley partial dislocations. By contrast, available literature data evidence that the degradation effect on 3C-SiC mechanical properties caused by comparable or even smaller point defect concentrations is many times stronger. Such a difference is likely due to extended defects only disturbing the lattice over a wider region without removing atoms. Therefore, covalent Si-C bonds are distorted rather than broken, which implies distributed elastic softening with only a moderate reduction in the values of the elastic constants. Use of the analytic bond-order potential for calculating 3C-SiC elastic constants and moduli provides generally better agreement with respective literature values than use of the Vashishta potential. As the extended defects concentrations grow, the Vashishta potential yields a more linear decrease in 3C-SiC mechanical characteristics, while the mechanical properties predicted by ABOP exhibit a somewhat oscillatory decline which is consistent with a documented higher ABOP sensitivity to local effects such as strengthening interactions between periodically densely packed defects.

## Acknowledgement

Authors acknowledge support of their work by the budget project III-1-26 "Ultra-high-frequency and ultra-fast electronic, phonon, spin, and plasmon processes in semiconductors, multifunctional nanomaterials, and nanostructures based on them" of the National Academy of Sciences of Ukraine.

**Authors' contributions**

**S.S.:** methodology, investigation, formal analysis, writing – original draft; **A.S.:** conceptualization, validation, writing – review and editing.